\documentclass[prl,aps,amsmath,amssymb,amsfonts,floatfix,groupeaddress,showpacs,twocolumn]{revtex4-1}
\pdfoutput=1
\usepackage{graphics}
\usepackage{color}
\usepackage{tikz}
\usepackage{dcolumn}
\usepackage{nicefrac}
\usepackage{multirow}
\usepackage{hyperref}
\newcommand{\bq}{\mathbf{q}}

\begin{document}
\title{Competing orders in Na$_x$CoO$_2$ 
from strong correlations on a two-particle level}
\author{Lewin Boehnke}
\affiliation{I. Institut f{\"u}r Theoretische Physik, 
Universit{\"a}t Hamburg, D-20355 Hamburg, Germany}
\author{Frank Lechermann}
\affiliation{I. Institut f{\"u}r Theoretische Physik, 
Universit{\"a}t Hamburg, D-20355 Hamburg, Germany}

\begin{abstract}
Based on dynamical mean-field theory with a continuous-time quantum Monte-Carlo impurity solver, static as well as dynamic spin and charge susceptibilites for the phase diagram of the sodium cobaltate system Na$_x$CoO$_2$ are discussed. The approach includes important vertex contributions to the $\bq$-dependent two-particle response functions by means of a local approximation to the irreducible vertex function in the particle-hole channel. A single-band Hubbard model suffices to reveal several charge- and spin-instability tendencies in accordance with experiment, including the stabilization of an effective kagom{\'e} sublattice close to $x$=0.67, without invoking the doping-dependent Na-potential landscape. The in-plane antiferromagnetic-to-ferromagnetic crossover is additionally verified by means of the computed Korringa ratio. Moreover an intricate high-energy mode in the transverse spin susceptiblity is revealed, pointing towards a strong energy dependence of the effective intersite exchange.
\end{abstract}

\maketitle
The investigation of finite-temperature phase diagrams of realistic
strongly correlated systems is a quite formidable task due to the often tight 
competition between various low-energy ordering instabilities. In this respect
the quasi-twodimensional (2D) sodium cobaltate system Na$_x$CoO$_2$ serves as 
a notably challenging case~\cite{foo04,lan08}. 
Here $x$$\in$[0,1] nominally mediates between the Co$^{4+}$($3d^5$,$\,S$=1/2) and 
Co$^{3+}$($3d^6$,$\,S$=0) low-spin states. Thus the Na ions provide the 
electron doping for the nearly filled $t_{2g}$ states of the triangular CoO$_2$ layers 
up to the band-insulating limit $x$=1. Coulomb correlations with a Hubbard $U$ up to 
5 eV for the $t_{2g}$ manifold of bandwidth $W$$\sim$1.5 eV~\cite{sin00} are revealed 
from photoemission~\cite{has04}. Hence with $U$/$W$$\gg$1 the frustrated 
metallic system is definitely placed in the strongly correlated regime. 

\begin{figure}[b]
\includegraphics{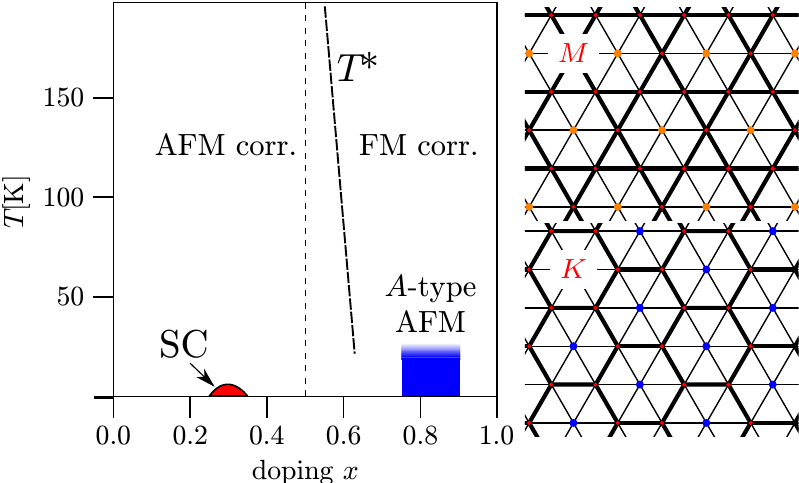}
\caption{(color online) Left: Sketched Na$_x$CoO$_2$ phase diagram, based on 
Ref.~\onlinecite{lan08}. Right: $M$- (top) and $K$- (bottom) point ordering on the 
triangular lattice.\label{nacoo2pd}}
\end{figure}

Various different electronic phases and regions for temperature $T$ vs. doping $x$
are displayed in the experimental sodium cobaltate phase diagram (see Fig.~\ref{nacoo2pd}).
For instance a superconducting dome ($T_{\rm c}$$\sim$4.5K) stabilized by 
intercalation with water close to $x$=0.3~\cite{tak03}. 
Pauli-like magnetic susceptibility is found in the range $x$$<$0.5~\cite{foo04} 
with evidence for 2D antiferromagnetic (AFM) correlations~\cite{fuj04,lan08}. 
For $x$$>$0.5 spin fluctuations and increased magnetic response show up for 
0.6$<$$x$$<$0.67, including the evolution to Curie-Weiss (CW) behavior~\cite{foo04} 
for 0.6$<$$x$$<$0.75, and the eventual onset of in-plane ferromagnetic (FM) 
order. The ordered magnetic structure in the doping range 0.75$<$$x$$<$0.9 with 
$T_{\rm N}$$\sim$19-27K~\cite{sug03,boo04,bay05,men05} is of 
A-type AFM for the FM CoO$_2$ layers. As the local spin-density approximation 
(LSDA) is not sufficient to account for the AFM-to-FM crossover with 
$x$~\cite{sin00}, explicit many-body approaches are needed~\cite{hae06,markot07,pie10}. 

Several theoretical works have dealt with the influence of the sodium arrangements on the 
electronic properties of Na$_x$CoO$_2$, both from the viewpoint of disordered sodium 
ions~\cite{markot07} as well as from orderings for certain dopings.~\cite{mer09,zho07,pei11}
However, wether such sodium patterns are due to sole (effective) single-particle potentials 
or mainly originating from many-body effects within the CoO$_2$ planes is still a matter of 
debate~\cite{zan04,hin08}.

In this letter, we report the fact that a large part of the electronic (spin and charge)
phase diagram of sodium cobaltate may be well described within a Hubbard model using
realistic dispersions, and without invoking the details of the sodium arrangement.
Thereby most of the observed crossovers and instabilities are truly driven by strong 
correlation effects and cannot be described within weak-coupling scenarios. The theoretical 
study is elucidating the two-particle correlations in the particle-hole channel computed 
within dynamical mean-field theory (DMFT) including vertex contributions (for a review see 
e.g.~\cite{geo96,mai05}). So far the latter have been neglected in cobaltate 
susceptibilities based on LSDA~\cite{joh042,kor07} and the fluctuation-exchange 
approximation~\cite{kur06,kor07}.   Our dynamical lattice susceptibilities 
allow to reveal details of the AFM-to-FM crossover with $T$ and of the intriguing 
charge-ordering tendencies, both in line with recent experimental data~\cite{lan08,all09}. 
Moreover, insight in the $(x,\bq)$-dependent spin excitations at finite frequency is 
provided.

Since we are mainly interested in the $x$$>$0.5 part of the phase diagram, the 
low-energy band dispersion of sodium cobaltate is described within an $a_{1g}$-like 
single-band approach, justified from photoemission~\cite{qia06_2} and 
Compton scattering~\cite{lav07} experiments. We primarily focus on the in-plane processes 
on the effective triangular Co lattice with tight-binding parameters up to 3rd 
nearest-neighbor (NN) hopping, i.e., ($t,t',t''$)=(-202, 35, 29)meV~\cite{ros03} for 
the 2D dispersion. Albeit intersite Coulomb interactions might play a role~\cite{pie10}, 
the canonical modeling was restricted to an on-site Coulomb interaction $U$=5 eV. 
Our calculations show that already therefrom 
substantial nonlocal correlations originate. The resulting Hubbard model on the 
triangular lattice is solved within DMFT for the local one-particle Green's function 
$G(\tau_{12})$=$-\langle T_{\tau}c(\tau_1)c^{\dagger}(\tau_2)\rangle$ with 
$\tau_{uv}$=$\tau_u$$-$$\tau_v$ and $T_{\tau}$ being the time-ordering operator. 
The DMFT problem is approached with the continuous-time quantum Monte Carlo 
methodology~\cite{rub05,wer06} in its hybridization-expansion 
flavor~\cite{wer06} as implemented in the \textsc{TRIQS} package.~\cite{fer11}
Additionally we implemented the computation of the impurity two-particle Green's 
function~\cite{boe11} $G^{(2)}(\tau_{12},\tau_{34},\tau_{14})$=$-\langle T_{\tau}c^{\dagger}(\tau_1)c(\tau_2)c^{\dagger}(\tau_3)c(\tau_4)\rangle$ to address explicit electron-electron correlations.
In the approximation of a purely local particle-hole irreducible vertex,
$G^{(2)}$ allows to determine also lattice susceptibilities.~\cite{zla90,geo96,mai05,boe11}
These susceptibilities, e.g. for spin $(s)$ and charge $(c)$, written as
\begin{multline}
\chi_{s/c}^{\hfill}(i\omega,\bq,T)=\\
T^2\sum_{\nu\nu'}\left(\tilde{\chi}^{(0)}_{s/c,\nu\nu'}(i\omega,\bq,T)
+v_{s/c,\nu\nu'}^{\hfill}(i\omega,\bq,T)\right)\,\,,\label{eq:gensus}
\end{multline}
where $\omega$ ($\nu$) marks bosonic (fermionic) Matsubara frequencies, consist of two
parts. Namely $\tilde{\chi}^{(0)}_{s/c,\nu\nu'}$ denotes the conventional (Lindhard-like) 
term, build up from the (renormalized) bubble part, which is mainly capable of detecting 
Fermi-surface driven instabilities close to $T$=$0$. On the contrary, the second part
$v_{s/c,\nu\nu'}$ (the vertex term) includes properly the energy dependence of the response 
behaviour due to strong local interactions in real space. It proves important for
revealing, e.g. magnetic instabilities at finite $T$ due to the resolution of the
two-particle correlations governed by an implicit inter-site exchange $J$. Note
that all numerics take advantage of the recently introduced Orthogonal Polynomial 
representation~\cite{boe11} of one- and two-particle Green's functions to provide the 
needed high accuracy and to eliminate artifacts often stemming from truncating the 
Fourier-transformed $G^{(2)}$ in Matsubara space.

Within the first Brillouin zone (BZ) of the triangular coordination with lattice constant 
$a$ the coherent $\Gamma$-point instability signals FM order in the case of 
$\chi_s^{\hfill}$ and phase separation for $\chi_c^{\hfill}$. Additionally important are 
here the instabilities at the the $K$- and $M$-point. The associated orderings give rise to 
distinct sublattice structures in real space (cf. Fig.~\ref{nacoo2pd}). The $M$-point 
ordering leads to a triangular and a kagom{\'e} sublattice with lattice constant 
$a_{\rm eff}$=$2a$, while the $K$-point ordering establishes a triangular and a honeycomb 
sublattice with $a_{\rm eff}$=$\sqrt{3}a$, respectively.

\begin{figure}[t]
\includegraphics{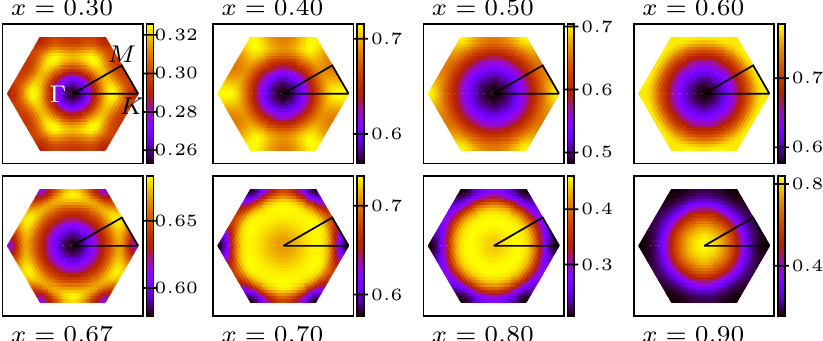}\\\vspace*{-0.2cm}
\rule{\linewidth}{1pt}\\\vspace*{0.1cm}
\noindent\includegraphics*{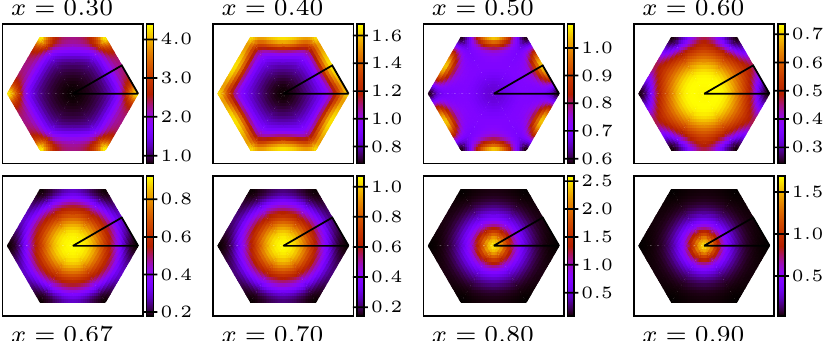}
\caption{(color online) Static in-plane charge (top) and spin (bottom) susceptibility 
$\chi(0,\bq,T)$ with doping at $T$=386K. 
\label{susplot}}
\end{figure}
We will first discuss the static ($\chi_{s/c}^{\hfill}(\omega$=$0,\bq,T)$) response
(read off from the zeroth bosonic Matsubara frequency), directly reflecting the system's
susceptibility to an order of the ($\bq$-resolved) type. The cobaltate intra-layer 
charge-susceptibility $\chi_c^{\hfill}(0,\bq,T)$ shows pronounced features in $\bq$ space 
with doping $x$ (see Fig~\ref{susplot}). Close to $x$=0.3 our single-band modeling leads to 
increased intensity inside the BZ, pointing towards longer-range charge-modulation 
(e.g. 3$\times3$, etc.) tendencies in real space. That Na$_{\nicefrac{1}{3}}$CoO$_2$ 
is indeed prone to such 120$^{\circ}$-like instabilities has been experimentally suggested 
by Qian {\sl et al.}~\cite{qia06}. Towards $x$=0.5 the susceptibility for short-range charge 
modulation grows in $\chi_c^{\hfill}$, displaying a diffuse high-intensity distribution at 
the BZ edge with a maximum at the $K$-point for $x$=0.5. No detailed conclusive result on 
the degree and type of charge ordering for the latter composition is known from experiment, 
however chain-like charge disproportionation that breaks the triangular symmetry is 
verified~\cite{hua04,nin08}. The present single-site approach cannot stabilize such 
symmetry-breakings, but an pronounced $\chi_c^{\hfill}$ at the $K$-point at least inherits 
some stripe-like separation of the two involved sublattices. Near $x$=0.67, the 
$\chi_c^{\hfill}$ maximum has shifted to the $M$-point, in line with the detection of an 
effective kagom{\'e} lattice from nuclear magnetic resonance (NMR) experiments~\cite{all09}.
For even higher doping, this $\bq$-dependent structuring transmutes into a $\Gamma$-point 
maximum, pointing towards known phase-separating tendencies.~\cite{lee06} 
Figure~\ref{susplot} also exhibits the $x$-dependent intra-layer spin susceptibility, 
starting with strong AFM peaks at $x$=0.3 due to $K$-point correlations. With 
reduced intensity these shift to the $M$-point at $x$=0.5, consistent with different 
types of spin and charge orderings at this doping level~\cite{nin08}. For $x$$>$$0.5$, 
$\chi_s^{\hfill}(\bq,T,0)$ first develops broad intensity over the full BZ, before forming a 
pronounced peak at the $\Gamma$-point above $x$$\sim$0.6. Thus the experimentally observed 
in-plane AFM-to-FM crossover in the spin response is reproduced.

\begin{figure}[t]
\includegraphics{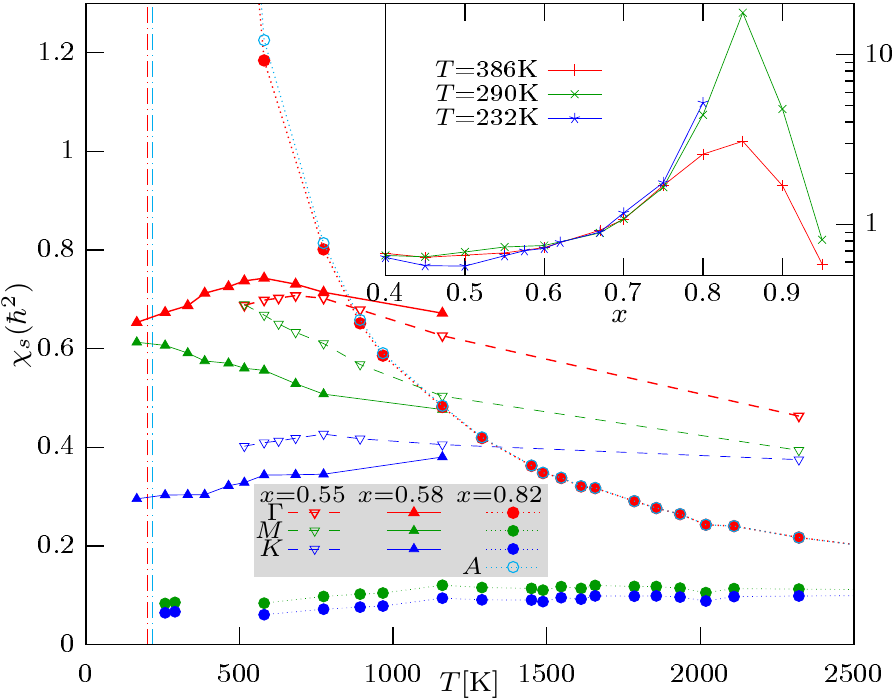}
\caption{(color online) Temperature dependence of the spin susceptibility at the $\Gamma$, $M$, and $K$-point for $x$=0.55, 0.58, 0.82. For the latter, values at the $A$-point are also included. Vertical lines indicate extrapolated transition temperatures for $\Gamma$ (FM) spin-ordering and $A$ (A-type AFM) spin-ordering respectively for $x$=0.82. The inset shows the doping dependence of the uniform ($\bq$=$\Gamma$) in-plane spin susceptibility $\chi_s^{\hfill}$ for various $T$. Note the largely increased magnitude of $\chi_s^{\hfill}$ for $x$$\ge$$0.75$ in this logscale plot.
\label{doptd}}
\end{figure}
Lang {\sl et al.}~\cite{lan08} revealed from the Na NMR that this crossover is $T$-dependent with $x$, resulting in an energy scale $T^*$ below which AFM correlations are favored (compare Fig.~\ref{nacoo2pd}). The slope $\partial T^*/\partial x$ turns out negative, in line with the general argument that FM correlations are most often favored at elevated $T$ because of the entropy gain via increased transverse spin fluctuations. In this respect, Fig.~\ref{doptd} shows the $(x,T,\bq)$ dependence of the computed $\chi_s^{\hfill}$. For $x$=0.55, 0.58 a maximum in the $\Gamma$-point susceptibility is revealed, which has been interpreted by Lang et al.~\cite{lan08} as the criterion for a change in the correlation characteristics, thereby defining the $T^*$-line. While the temperature scale exceeds the experimental value in the present mean-field formalism, the qualitatively correct doping behavior of the $T^*$-line is obtained.

Beyond the experimental findings our calculations allow to further investigate the nature of the magnetic crossover. Fig.~\ref{doptd} reveals that at lower $T$ and $x$ closer to $x$=0.5 the susceptibility at $\Gamma$ is ousted by the one at $M$, while $\chi_s^{\hfill}$ at $K$ is mostly dispensable. The $M$ susceptibility can be understood due to the proximity of the striped order at $x$=0.5,\cite{foo04,zan04,gas06} which is however not realized until much lower temperatures.

The inset of Fig.~\ref{doptd} follows the $T$-dependent $\Gamma$-point susceptibility through a
vast doping range. Note the subtle resolution around $x$=0.5 as well as the large exaggeration 
especially for lower temperatures in the experimentally verified in-plane FM region. 
The main panel of Fig.~\ref{doptd} additionally shows for $x$=0.82 the spin susceptibility
at the $A$-point (i.e., at $k_z$=(0,0,1/2) in the BZ), which denotes the 
A-type AFM order. While $\Gamma$ and $A$ show CW behavior, the extrapolated transition temperature 
however is $\sim$7$\%$ higher at $A$ than at $\Gamma$, verifying the experimental findings of 
A-type order~\cite{sug03,boo04,bay05,men05}. In the temperature scan we additionally introduced 
a nearest-layer inter-plane hopping $t_{\perp}$=13 meV~\cite{fou73,bay05,pie10}, however the 
previous in-plane results are qualitatively not affected by this model extension.
Due to known charge disproportionation the inclusion of long-range Coulomb interactions, e.g., via an inter-site $V$~\cite{pie10,pei11}, seems reasonable. This was abandoned in the present single-site DMFT approach, resulting generally in reduced charge response. Without $V$, charge fluctuations are substantially suppressed for large $U$/$W$, while the inter-site spin fluctuations are still strong due to superexchange.
\begin{figure}[b]
\includegraphics{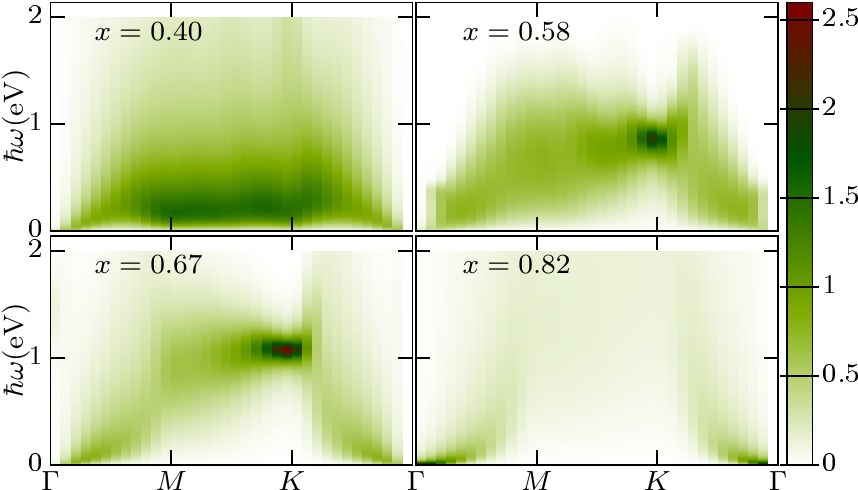}
\caption{(color online) Imaginary part of the dynamical spin-susceptibility for 
selected dopings for selected dopings.}
\label{fig:dynsus}
\end{figure}
Aside from the static response, our method allows also access to the dynamic regime. 
Figure~\ref{fig:dynsus} shows the dynamical transverse spin susceptibilitiy for 
selected $x$. Note the broad $\bq$-dependence and small excitation energy in the 
low-doping regime. In contrast, the FM correlations near $x$=0.82 are reflected by strong 
paramagnon-like gapless excitation at $\Gamma$ combined with very little weight and rather 
high excitation energies at AFM wave-vectors. Interestingly, a comparably strong and sharp 
$K$-type high-energy excitation ($\sim$1 eV )for larger $x$ below the onset of in-plane FM 
order is revealed. Its amplitude is strongest at $x$=0.67 while its energy increases with 
$x$ and its worthwhile to note that the mode is \emph{neither} visible when neglecting vertex 
contributions, \emph{nor} in a plain triangular Hubbard model with NN hopping only. Thus it 
reflects a strong energy dependence of the inter-site exchange coupling 
$J$=$J(x,{\bf q},\omega)$, that obviously changes character for $x$$\sim$0.67 with
$\bq$ and $\omega$. The predicted high-energy feature could be probed experimentally 
and also studied in time-dependent measurements. We propose the use of modern laser-pulse 
techniques~\cite{kim06} to address this problem.

Experimentally, the evidence for significant $\bq$$\neq$0 fluctuations is drawn~\cite{all08,lan08} from the Korringa ratio~\cite{kor50,mor63,yus09}\footnote{Note that we use a slightly different diffinition of $\chi_s$ than Ref.~\cite{yus09}.}
\begin{align}
\mathcal{K}^T_x&=\frac{\hbar}{4\pi k_B}\left(\frac{\gamma_e}{\gamma_N}\right)^2\frac{1}{T_1TK_S^2}\notag\\
\frac{1}{T_1T}&=\lim_{\omega\to0}\frac{2k_B}{\hbar^2}\sum_\bq|A(\bq)|^2\frac{\Im\chi^{-+}_{s}(\omega,\bq,T)}{\omega}\label{eqn:kor}\\
K_S&=\frac{|A(\mathbf{0})|\gamma_e\Re\chi^{-+}_s(0,\mathbf{0},T)}{\gamma_N\hbar^2}\notag
\end{align}
where $1/T_1$ is the nuclear relaxation rate, $K_S$ the NMR field shift, $\gamma_e$ ($\gamma_N$) the electronic (nuclear) gyromagnetic ratio and $A(\bq)$ the hyperfine coupling. Roughly speaking, ${\cal K}$$>$1 signals AFM correlations, ${\cal K}$$<$1 points to FM tendencies in $\chi_s^{\hfill}$ and the term ``Korringa behavior'' generally denotes the regime ${\cal K}(T)$$\sim$1. In single-atom unit cells, $A(\bq)$ becomes $\bq$-independent.
\begin{figure}[t]
\includegraphics{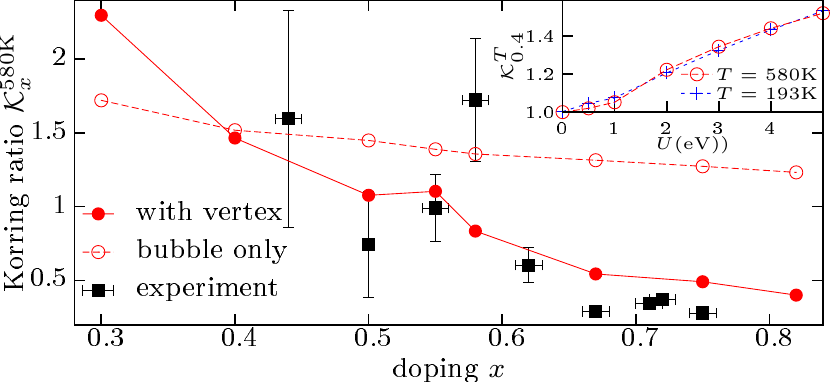}
\caption{(color online) Korringa ratio versus doping for $T$=580 K. The experimental data, 
extracted from Ref.~\onlinecite{all08} and \onlinecite{lan08}, was
obtained for lower temperatures. The inset shows the evolution of the bubble-diagram 
contribution from the non-interacting ($U$=0) to the fully-interacting ($U$=5 eV) case.}
\label{fig:kor}
\end{figure}

Note that especially $1/T_1$ is numerically expensive, as it requires to calculate 
$\chi^{-+}_s$ on many Matsubara frequencies with subsequent analytical continuation to 
the real frequency axis for contributions beyond the bubble diagram. Figure~\ref{fig:kor} 
finally shows the AFM-to-FM correlation crossover captured by the Korringa-ratio over a 
wide doping range.
The overall agreement with experiment is conclusive. Relevant deviations in the low-doping regime probably originate from the smaller temperatures studied in experiment. The difference at $x$=0.58 might be of the same origin, but since charge ordering occours for $x$$>$0.5 which was not included explicitly here, neglecting the $\bq$-dependence of $A(\bq)$ might be also questionable.\footnote{A further reason could be an additional peak in the corresponding measurement at this precise doping\cite{lan08}, which might or might not be of electronic nature, possibly influencing the Korringa ratio.} One can see that the bubble-only calculation yields a nearly flat Korringa ratio with doping, thus fails completely in explaining the 
experimental findings. In particular it does not reflect the strong FM correlations for high 
doping. This further proves the importance of strong correlations on the two-particle level,
asking for substantial vertex contributions.~\cite{yus09} Note that the recently suggested 
lower-energy effective kagom{\'e} model~\cite{pei11} including the affect of charge 
ordering is not contradicting the present modeling. Since here the effective kagom{\'e} 
lattice naturally shows up and also the key properties of the spin degrees of freedom 
seem well described on the original triangular lattice.

In summary, the DMFT computation of two-particle observables 
including vertex contributions based on a realistic single-band Hubbard modeling 
for Na$_x$CoO$_2$ leads to a faithful phase-diagram examination at larger $x$, including the 
kagom{\'e}-like charge-ordering tendency for $x$$\sim$0.67 and the in-plane
AFM-to-FM crossover associated with a temperature scale $T^*$. Thus it appears that many
generic cobaltate features are already governed by a canonical correlated
model, without invoking the details of the doping-dependent sodium-potential 
landscape or the inclusion of multi-orbital processes. Of course, future work has to 
concentrate on quantifying further details of the various competing instabilities 
(and their mutual couplings) within extended model considerations. Beyond equilibrium
physics, we predict a strong energy dependence of the effective inter-site exchange 
resulting in an $K$-type high-energy mode around $x$=0.67, which could be probed in 
experimental studies.
\acknowledgements
The authors are indebted to O.~Parcollet and M.~Ferrero for useful discussions on the vertex implementation and calculation. We also like to thank A.~Georges, H.~Hafermann
A.~I.~Lichtenstein, O.~Peil and C.~Piefke for valuable comments. This work was supported by the SPP 1386 and the FOR 1346 of the DFG. Computations were performed at the North-German Supercomputing Alliance (HLRN) and the regional computing center (RRZ) of the University of Hamburg.
\bibliography{bibextra}
\end{document}